\documentstyle[epsf,twocolumn,seceq]{jpsj}

\def\runtitle{Thermal Properties of Heavy Fermion Compound YbP}
\def\runauthor{Naoyuki {\sc TATEIWA} {\it et al}}

\title
{
Thermal Properties of Heavy Fermion Compound YbP
}

\author
{ 
Naoyuki {\sc TATEIWA}\footnote{E-mail:tateiwa@djebel.mp.es.osaka-u.ac.jp}, Tatsuo C {\sc KOBAYASHI}, Kiichi {\sc AMAYA}$^{1}$, Dexin {\sc LI}$^{2}$,\\Yoshinobu {\sc SHIOKAWA}$^{2}$ and Takashi {\sc SUZUKI}$^{3}$
}

\inst
{

 Research Center for Materials Science at Extreme Conditions, Osaka University, Toyonaka, Osaka 560-8531, Japan\\
$^1$Department of Physical Science, Grasuate School of Engineering Science, Osaka University, Toyonaka, Osaka 560-8531, Japan\\
$^2$ Oarai Branch, Institute for Materials Research, Tohoku University, Oarai, Ibaraki, 311-1313, Japan\\
$^3$ Physics Department, Graduate School of Science, Tohoku University,Sendai 980-8545, Japan\\

}

\recdate
{
May 10, 2001
}

\abst
{
Low-temperature specific heat and its field-dependence up to 16 T was measured in a stoichiometric single crystal of YbP. A sharp peak was observed at {\it T}$_{\rm N}$ = 0.53 K in zero magnetic field. Application of external field seems to induce a new magnetic phase above 11 T. The field dependence of the transition temperature in the high-field phase is different from that of the low field phase. The linear coefficient of the electronic specific heat is estimated as 120 mJ/mole K$^{2}$ from low temperature specfic heat, suggesting heavy Fermion state in YbP.
}

\kword
{
YbP, magnetic phase diagram, heavy-Fermion state
}

\begin{document}
\sloppy
\maketitle

\section{Introduction}

  The Yb monopnictides YbX$_{\rm p}$ (X$_{\rm p}$ = N, P, As and Sb) with the cubic NaCl-type crystal structure are semi-metallic compounds with an extremely low carrier concentration.~\cite{rf:1,rf:2}  The Yb ion in YbX$_{\rm p}$  is predominantly trivalent with one 4{\it f} hole, in which the spin-orbit split $J = 7/2$ is the ground state configuration and it further splits into ${\mit\Gamma}_6$, ${\mit\Gamma}_8$, and ${\mit\Gamma}_7$ states in the cubic crystal field with ${\mit\Gamma}_6$ doublet as the ground state.~\cite{rf:3,rf:4,rf:5} The band calculation indicates that the bottom of the conduction band, formed mainly by 5{\it d} (Yb), is at each X point, slightly overlapping with the top of the valence band to make semimetals with a low carrier concentration of the order of 0.01 per Yb$^{3+}$ ion.~\cite{rf:6,rf:7} The Yb monopnictides have attracted a lot of attention in the last decade because of their various puzzling properties. At around 0.5 K, the stoichiometric compounds YbN and YbAs show antiferromagnetic fcc type-III order, and nonstoichiomatric YbP$_{0.84}$ undergoes a magnetic ordering corresponding to fcc type-II antiferromagnetism.~\cite{rf:8,rf:9,rf:10} The ordered magnetic moments in YbN, YbP and YbAs are much smaller than the value of 1.33 $\mu_{\rm B}$ expected from the crystalline-electric field (CEF) ground state doublet ${\mit\Gamma}_6$. The specific heat measurement shows a broad peak around 5 K and the entropy associated with the antiferromagnetic ordering are 20-30 \% of ${\it R}$ ln2 in YbN, YbP and YbAs. Moreover, the coefficient of electronic specific heat $\gamma$  was estimated to be 270 mJ/moleK$^2$ for YbAs.~\cite{rf:11} From these results, Yb monopnictides have been considered to be heavy-Fermion compounds with the Kondo effect. However, the -ln${\it T}$ behavior characteristic of Kondo compounds is not detected in the electrical resistivity of YbAs and YbP. There is not a theoretical model that could give a comprehensive interpretation of all the experimental results.  Clearly, the anomalous physical properties in YbXp should be studied in more detail. In particular, experimental works on high quality single crystals are necessary.

  It is very difficult to grow single crystals of YbX$_{\rm p}$ due to the high melting point and a high vapor pressure. So far, single crystals of YbX$_{\rm p}$ were prepared only for YbAs and YbP, but the YbP sample was a nonstoichiometric one, ~\cite{rf:12} and thus intrinsic physical property, in particular transport properties, was not reported for YbN, YbP and YbSb until now. Recently, Li {\it et al.} has succeeded in growing the stoichiometric single crystal of YbP. Their ex  perimental results revealed the smaller residual resistivity $RR$ = $\rho (T\rightarrow 0{\rm \, K}) = 6.9 \,\mu\Omega{\rm cm}$, the larger residual resistance ratio $RRR$ = $\rho (T=300{\rm \, K})/\rho (T\rightarrow 0{\rm \, K}) = 8.5$, and the larger positive magnetoresistance $MR$=$[\rho (H)-\rho (0)]/\rho (0)=1.23$ (at $H$ =10 T and $T$ =4.2 K).  The corresponding values for nonstoichiometric YbP$_{0.84}$ are $RR$=17.2 $\,\mu\Omega{\rm cm}$, $RRR$ = 3 and $MR$ = 0.08. To our knowledge, this YbP single crystal is the best sample at present. In this paper, we present the result of specific heat measurement performed on this YbP single crystal under magnetic field up to 16 T applied along $\langle 100 \rangle$ direction. A large coefficient of electric specific heat and a new magnetic phase under high magnetic field were detected in YbP. 
\begin{figure}
 \begin{center}
  \epsfxsize=8.5cm
  \epsfbox{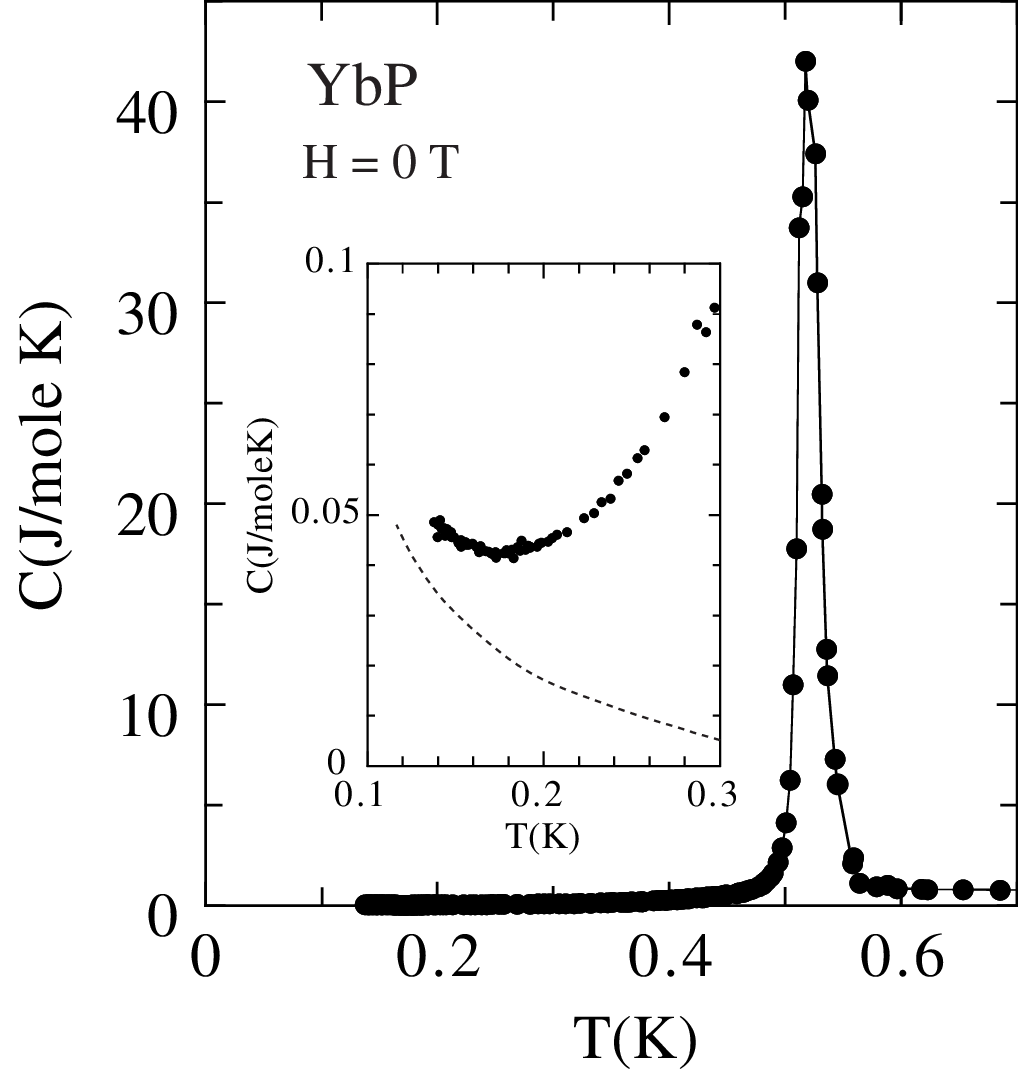}
 \end{center}
\caption{Temperature dependence of the specific heat of YbP in zero field. A sharp peak near $T$ = 0.5 K indicates the antiferromagnetic phase transition. The inset shows the low temperature part. The dotted line indicates the nuclear contribution of the specific heat which is estimated from the hyperfine field derived from the measurement of $^{170}{\rm Yb}$ M\"{o}ssbauer effect. }
\label{fig:1}
\end{figure}
\section{Experimental}

 The single crystal sample used in this study was grown by the mineralization method in a tungsten crucible. The detailed description of the sample preparation is given in ref. 13. The powder sample was first prepared by pre-reacting starting elements with a composition of Yb:P = 1:1.1 in a closed quartz ampoule. The elements were heated for 2 weeks up to 380 $^{\circ}$C, and annealed for 1 week at 700 $^{\circ}$C. The neutron diffraction experiments were performed on this powder sample by Keller {\it et al.}~\cite{rf:9}, and revealed the average phosphorus content of $x$ = 0.99(1) per formula  YbX$_{x}$ indicating almost ideal stoichiometry. The powder material of YbP was then pressed into a hard pellet and sealed in tungsten crucible. Finally, the crucible was heated up to above 2500 $^{\circ}$C and kept at this temperature for 72 hours. A single crystal of about $5 \times 5 \times 5$ mm$^3$ was obtained in this way. X-ray-diffraction patterns showed a NaCl-type single phase for this sample with the lattice parameter value of 5.542 {\AA} at room temperature. Chemical analysis yielded the value of 1:1.00 $\pm$ 0.01 for the atomic ratio between Yb and P consistent with the neutron-diffraction results mentioned above. We measured the specific heat by means of a conventional adiabatic heat pulse method in the temperature range of 0.13 - 12 K generating by a $^3$He-$^4$He dilution refrigerator. An external magnetic field up to 16 T was applied using a superconducting magnet. A RuO$_2$ resistance thermometer is located in the compensated field in order to avoid the field dependence of the resistance.

\section{Results and Discussion}
The temperature dependence of the specific heat $C(T)$ in zero field is shown in Fig. 1. The peak at 0.53 K is due to the antiferromagnetic phase transition.
\begin{figure}
 \begin{center}
  \epsfxsize=8.5cm
  \epsfbox{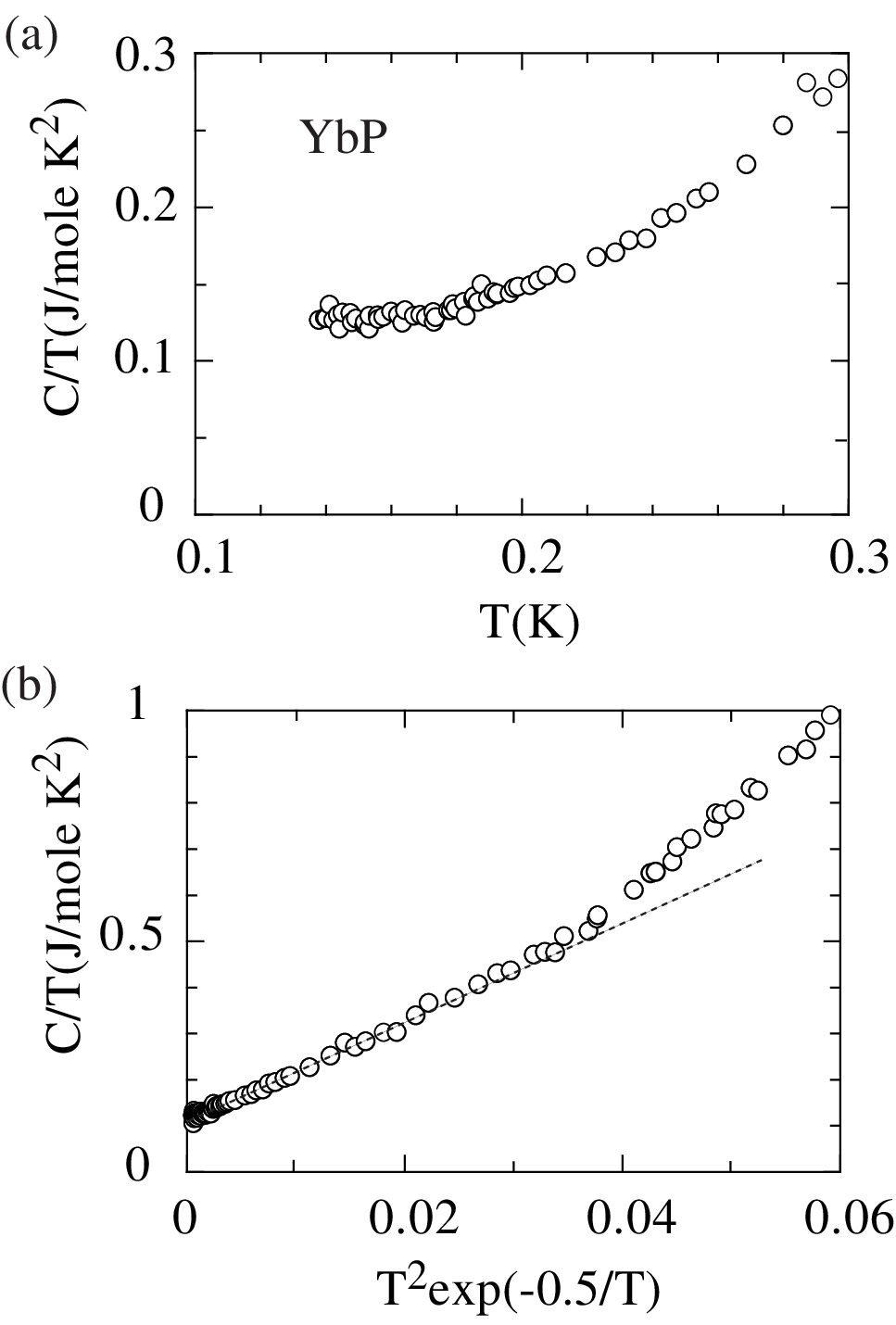}
 \end{center}
\caption{(a) $C/T$ vs. $T$ plot of YbP. $C/T$ shows an almost constant value of about 130 mJ/moleK$^2$ below 170 mK. (b)  $C/T$ vs. $T^2$exp(-0.5/$T$) plot of YbP. The dashed line represents "least-squares" fit using Eq.(1) in the temperature range from 138 mK and 350 mK.}
\label{fig:2}
\end{figure}
 The peak is sharp and the peak value is three times larger than that in the previous result, indicating a good quality of our single crystal sample.~\cite{rf:10}The increase of the specific heat below 170 mK is attributed to the nuclear Zeeman effect of Yb nuclei suffering an exchange field in the antiferromagnetic state. 

 The specific heat of YbP contains a phonon part $C_{\rm p}$(proportional to $T^3$, which is important only at higher temperature and thus is neglected in the following analysis), a nuclear part  $C_{\rm n}$ an electronic part $C_{\rm e}$ and a magnetic part $C_{\rm m}$. We estimated the nuclear contribution of the specific heat from the hyperfine field ( $H_{\rm hf}$ = 93 T ) derived from the measurements of the $^{170}$Yb M\"{o}sbauer effect as denoted in dotted line in Fig. 1.~\cite{rf:14} By subtracting the nuclear contribution from the total specific heat, the electronic specific heat $C(T)$ (= $C_{\rm e}$+$C_{\rm m}$ ) was obtained and plotted in Fig. 2(a). $C/T$ seems to approach constant value around 130 mJ/moleK$^2$ below 170 mK.
@
\begin{figure}
 \begin{center}
  \epsfxsize=8.5cm
  \epsfbox{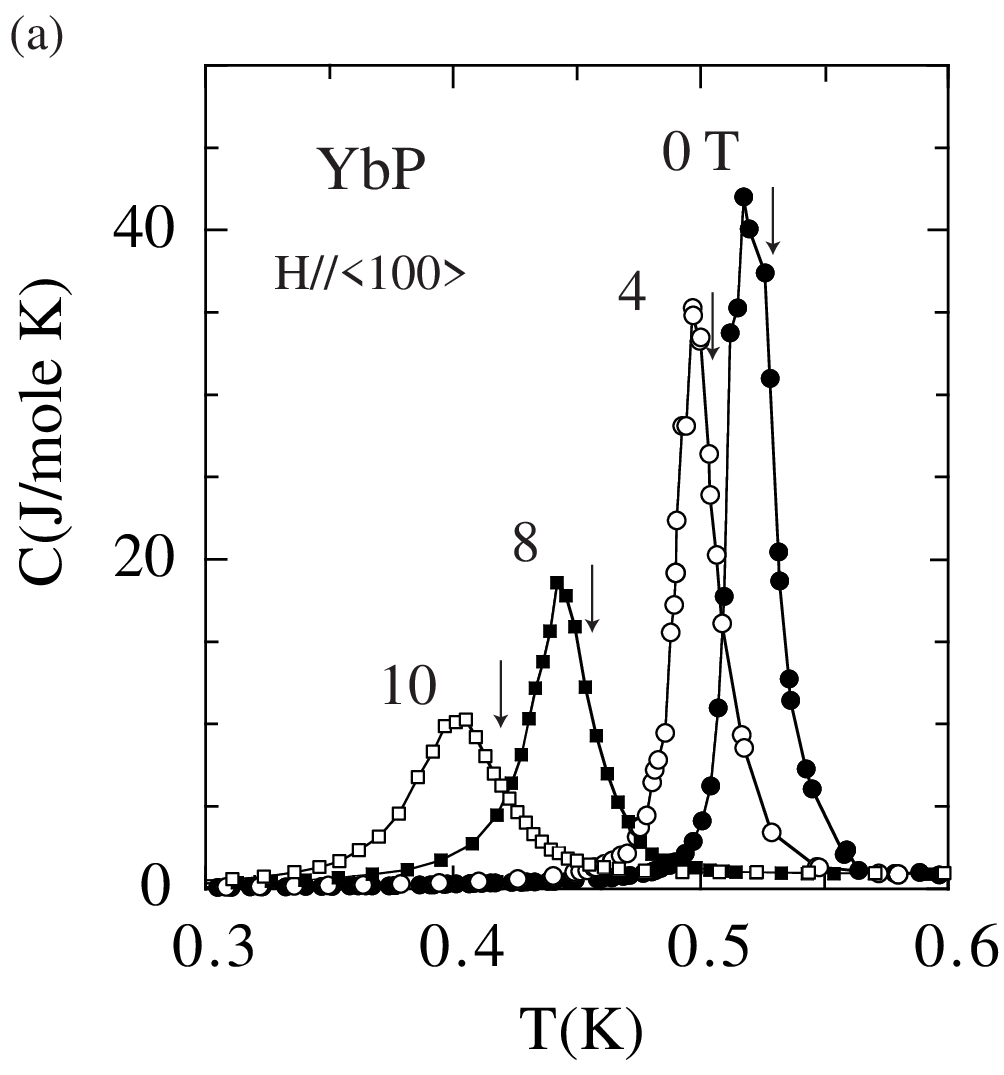}
\epsfxsize=8.5cm
  \epsfbox{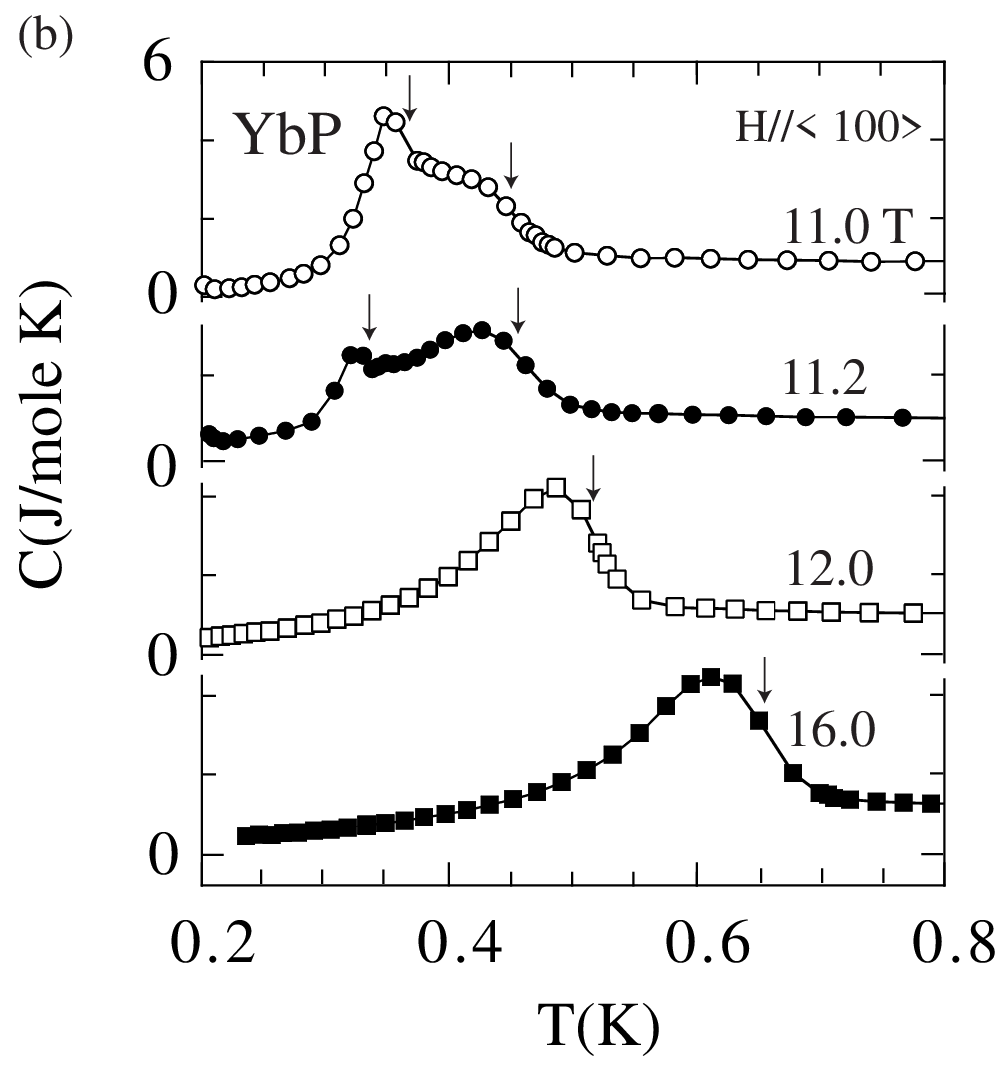}
 \end{center}
\caption{Specific heat of YbP in magnetic ifelds up to 16 T. $C(T)$ shows a single discontinuity at $T_{\rm N}$ below 10T (a). A two peak structure appear at 11.0 T and 11.2 T (b). The transition temperatures are denoted by arrows.}
\label{fig:3}
\end{figure}

Next we consider the contribution from the antiferromagnetic spin wave excitation to the specific heat $C_{\rm m}$. We could not fit the experimental result if we assumed that antiferromagnetic spin wave dispersion does not have a gap and $C_{\rm m}$ is expressed as $C_{\rm m} \sim T^3 $. Thus, it is suggested that the antiferromagnetic spin wave excitation of YbP has a gap similar to that of YbAs. In YbAs, D\"{o}nni {\it et al.} have performed an inelastic neutron scattering experiment and reported that one of two magnon branches has a $\it k$-linear dispersion near the magnetic zone center with an energy gap of 0.08 meV.~\cite{rf:15}
\begin{figure}
 \begin{center}
  \epsfxsize=8.5cm
  \epsfbox{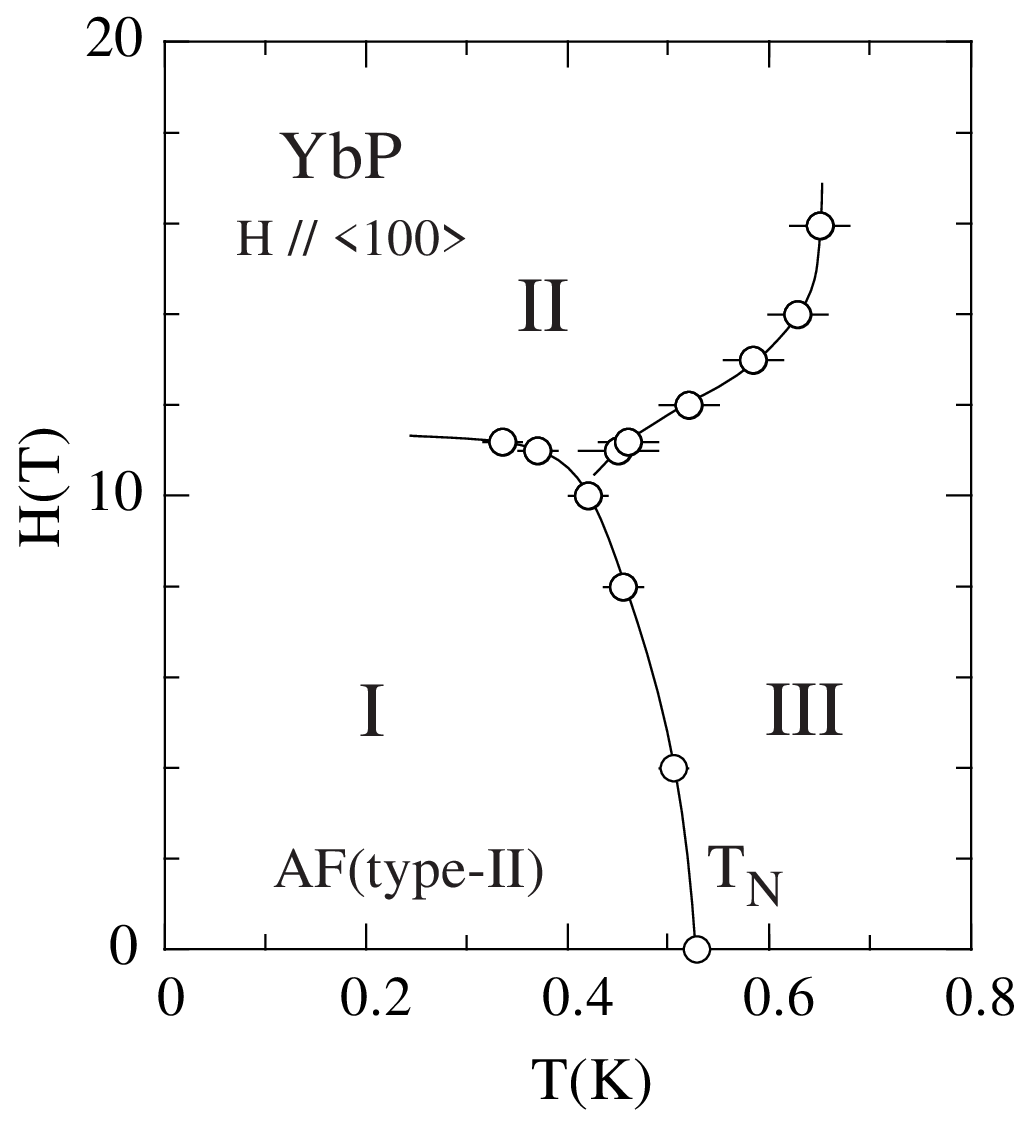}
 \end{center}
\caption{Magnetic phase diagram of YbP for the magnetic field along $\langle100 \rangle$ direction obtained from the specific heat measurements}
\label{fig:4}
\end{figure} 
 We fit the measured specific heat data using the expression
  \[C= C_{\rm e}+C_{\rm m} = \gamma T + {\mit\delta} T^ 3 {\rm exp}(-{\mit\Delta}/T)\,\,\,\,\,\,(1)\]
in the temperature range from 138 mK to 350 mK.  The first term indicates the electronic specific heat $C_{\rm e}$. The second term indicates the antiferromagnetic magnon contribution to the specific heat $C_{\rm m}$ and we use a same expression that Sakon {\it et al.} used in the analysis of the specific heat of YbAs.~\cite{rf:11} The energy gap ${\mit\Delta}$  and the coefficient of the electronic specific heat $\gamma$  are estimated to be 0.5 K and 115 mJ/moleK$^2$ mole, respectively. This $\gamma$  value is compatible with the constant $C/T$ value observed below 170 mK. 

 We calculated $C_{\rm m}$ with eq. (1) using obtained parameters and found that $C_{\rm m}$is only 6 \% of the specific heat $C(T)$ (= $C_{\rm e}+C_{\rm m}$ ) at 150 mK. Thus, about 90 \% of the constant $C/T$ observed below 170 mK comes from the electronic specific heat $C_{\rm e}/T$. From these results, we conclude that the $\gamma$  value is about 120 mJ/moleK$^2$ in YbP. Note that a large $\gamma$ value of 270 mJ/moleK$^2$ was also reported for YbAs. ~\cite{rf:11}

The $\gamma$ values of YbP and YbAs are much larger than those of normal metals, indicating the formation of heavy fermion state in both systems. The anomalous physical properties in Yb monopnictides are usually interpreted in terms of the competition between the Kondo effect and RKKY interaction. In this mechanism, the Kondo effect was due to a large mixing between 4{\it f} ${\mit\Gamma}_6$ hole state of the Yb$^{3+}$ ion and the occupied ${\it p}-{\mit\Gamma}_6$ valence-hole states. Thus, the existence of ${\it p}-{\mit\Gamma}_6$ holes in YbXp is the prerequisite for the Kondo effect. 
\begin{figure}
 \begin{center}
  \epsfxsize=8.5cm
  \epsfbox{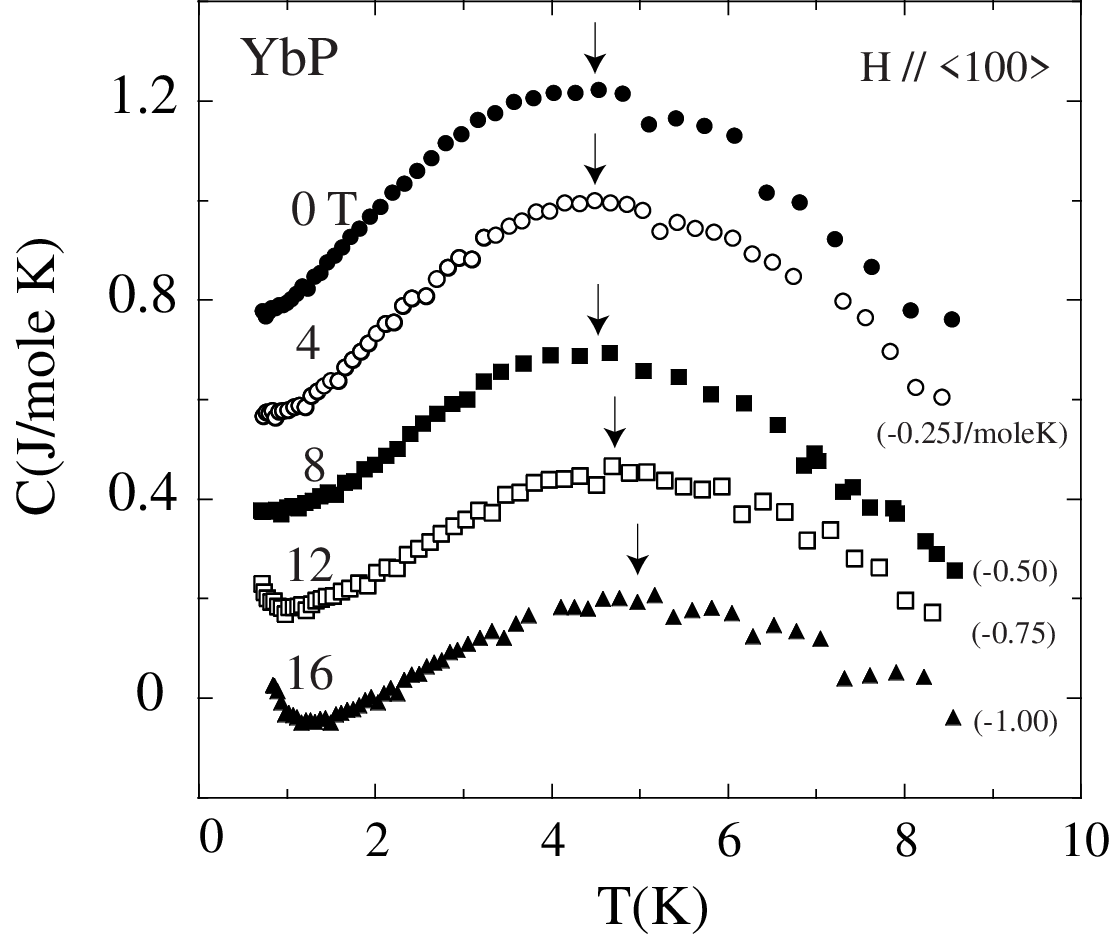}
 \end{center}
\caption{Temperature dependence of the specific heat of YbP at higher temperatures in the magnetic fields up to 16 T. The curves in higher fields are sifted downwards for the sake of clarity.}
\label{fig:5}
\end{figure}

Usually, for a metallic heavy Fermion compound, the concentration of conduction electrons is nearly the same order of magnetic ions. However, the problem is that the total hole carrier numbers in YbAs and YbP are the order of 0.01 per Yb atom. Moreover ${\it p}-{\mit\Gamma}_6$ hole was not observed in the dHvA experiment on YbAs.~\cite{rf:16} Sakon {\it et al.} and Hashi {\it et al.} proposed a conjecture that heavy Fermion state in Yb monopnictides may be realized by antiferromangetic correlations between the localized moments of the Yb ion due to the exchange interaction.~\cite{rf:11, rf:17, rf:18} In this mechanism, the magnitude of the localized moments is reduced by this antiferromagnetic correlations and the Kondo type screening of the localized moments just help the mutual reduction of the localized moments. On the different point of view, a charge dipolar model was recently proposed and used to explain the experimental results of Yb monopnictides by Kasuya {\it et al.}~\cite{rf:19} At present, it is impossible to discuss the validity of above scenarios within the present experimental results because these theoretical predictions are limited within qualitative considerations. In order to clarify the heavy fermion state in Yb monopnictides, further theoretical investigations and quantiative comparison with the theories and experimental results are needed.

   Fig.3 (a) and (b) show the temperature dependence of the specific heat of YbP under magnetic fields between $0 < H <10 {\, \rm T}$ and $11 < H <16 {\, \rm T}$, respectively, applied along the $\langle100 \rangle$ direction. Below 10 T, a single peak appears in $C(T)$ at the N\'{e}el temperature $T_{\rm N}$ . With increasing $H$, the peak loses intensity and the position shifts to lower temperature.  It is noted that $C(T)$ curve shows two peaks in the external fields of 11.0 T and 11.2 T as illustrated in Fig. 3 (b), indicating that there clearly exists an additional phase transition in the ordered phase. We defined the transition temperature such that the entropy is conserved, that is, entropy balance. In this way, the N\'{e}el temperature of our YbP sample is estimated to be 530 mK at zero magnetic field.
  \begin{figure}
 \begin{center}
  \epsfxsize=8.5cm
  \epsfbox{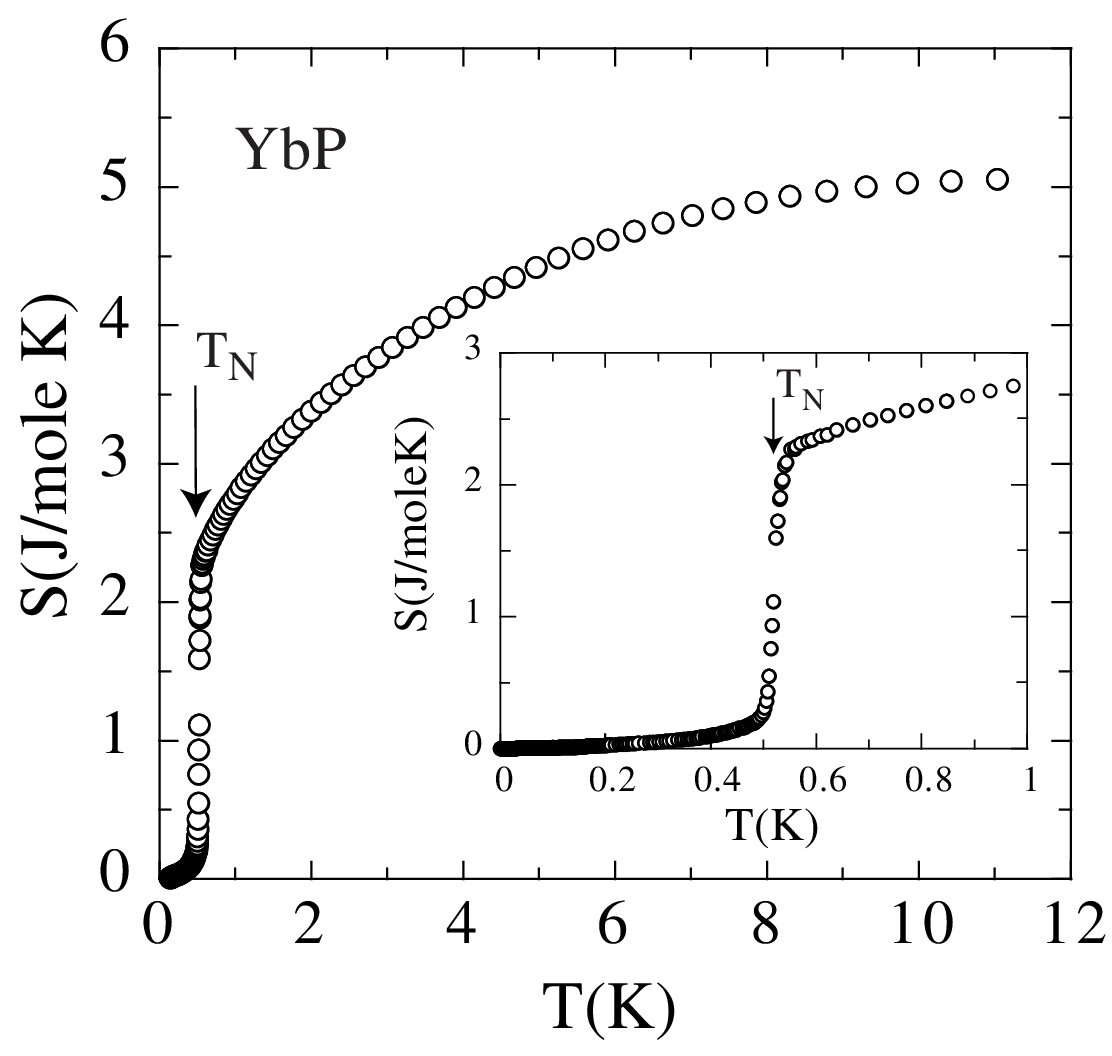}
 \end{center}
\caption{Temperature dependence of the entropy of YbP in zero field.}
\label{fig:6}
\end{figure}
   This value is smaller than {\it T}$_{\rm N}$ = 660 mK determined by neutron diffraction experiments for the prereacted powder sample.~\cite{rf:9} It may be attributed to either sample dependence of difference in the definition of transition temperature. With these transition temperatures, we have determined a magnetic phase diagram as shown in Fig. 4. Here, the phases are referred to as I, II and III. The phase I is the antiferromagnetic ordered state which is previously reported. The phase II is revealed for the first time in this measurement. The phase III is a paramagnetic phase. As the field strength increases, the transition temperature from the phase I to III shifts to lower temperature. On the other hand, the transition temperature from II to III increases with the magnetic field. 
   
 In Ce monopnictides such as CeSb, CeP and CeAs, complicated magnetic phase diagrams or multi step meta-magnetic transitions were observed.~\cite{rf:20} Neutron scattering experiments revealed that complicated magnetic structures appear in the magnetic field in Ce monopnictides.~\cite{rf:21, rf:22}  Many theoretical studies have been done in order to explain the unusual magnetic properties of Ce monopnictides.~\cite{rf:20} Note that the magnetic phase diagram of YbP has some similarity with that of Ce monopnictide. It is interesting to perform neutron scattering experiment or magnetization measurement on YbP and compare magnetic properties with those of Ce monopnictides.

 Fig.5 shows the temperature dependence of the specific heat in high temperature region. For the sake of clarity, we omit the data below 0.8 K and the curves in the higher fields are shifted downwards by the values in parentheses. A broad maximum exists around 4 K in $C(T)$ curve, which is consistent with the previous reports.~\cite{rf:10, rf:23} This peak was considered to be a Kondo peak with $T_{\rm K}$ of about 5.7 K.~\cite{rf:23} Recently the results of the inelastic neutron scattering experiment showed that short range antiferromagnetic correlations in YbX$_{\rm p}$ begin to develop below about 20 K far above T$_{\rm N}$~\cite{rf:24}, and then the broad specific heat maximum may be related to this antiferromagnetic correlation. The position of the broad maximum shows no distinct change in lower fields and slightly sifts to higher temperatures with increasing the field above 8 T. This feature can not be explained in terms of antiferromagnetic correlations. We suggest that a short range ferromagnetic correlation is induced in the magnetic field above 10 T and this correlation may be related to the appearance of the phase II above 10 T. 

We have also estimated the entropy by integrating $C/T$ curve with respect to $T$ as shown in Fig. 6. To obtain the entropy $S$, we extract the nuclear part of the specific heat, $C_{\rm n}$ obtained from the M\"{o}ssbauer measurement as described above. We extrapolate $C/T$ curve below 138 mK to zero temperature using a relation $C = \gamma^{'} T$ with $\gamma^{'}$ = 130 mJ/moleK$^{2}$. The appearance of a sharp peak in $C(T)$ and sudden change in $S(T)$ at $T_{\rm N}$ suggest that a first-order phase transition occurs at this temperature. At $T_{\rm N}$, $S$ reaches 40 \% of $R$ ln2 = 5.76 J/moleK, the expected entropy of a ${\mit\Gamma}_6$ doublet ground state. This $S$ value is two times larger than that reported in Ref.10 in which the N\'{e}el temperature of YbP was determined to be 0.41 K. The sample quality may be responsible for this difference. In this work, we used a high quality single crystal sample that shows the higher N\'{e}el temperature (0.53 K) and a larger transition peak. Thus the estimation of the entropy in the present study is more reliable. The total entropy $R$ ln2 is recovered if the temperature is raised up to the temperature much above 10 K, that is, more than an order of magnitude above the magnetic ordering temperature. This feature resembles to that of the dense Kondo Cerium compounds and again suggests the formation of heavy Fermion state in YbP. 
 
\section{Conclusion}
 In conclusion, we have measured the low temperature specific heat of YbP up to 16 T using a stoichiometric single crystal sample. The $\gamma$ value and the magnetic phase diagram in YbP were determined. The large $\gamma$ value and the reduced entropy at the magnetic ordering temperature suggest the formation of heavy Fermion state in YbP. A new magnetic phase was observed under the high magnetic field of YbP and the magnetic phase diagram has some similarity comparing with that of Ce monopnictides. This suggests that some interesting magnetic phenomena similar to those observed for Ce monopnictides could also be expected for Yb monopnictides. 

%\vspace{1zw}
%{\bf Acknowledgement}
%
%This work was financially supported by CREST, Japan Science and Technology Corporation. 

\end{document}